\documentclass[aps,pra,10pt,tightenlines,notitlepage,superscriptaddress,amsmath,twocolumn,floatfix,eqsecnum]{revtex4-1}
\usepackage[dvipsnames]{xcolor}
\usepackage[colorlinks=true,bookmarks=false,linkcolor=NavyBlue,urlcolor=NavyBlue,citecolor=NavyBlue,breaklinks]{hyperref}
\usepackage{amsmath}
\usepackage{amssymb}
\usepackage{amsfonts}
\usepackage{graphicx}
\usepackage{braket}
\usepackage{mathtools}

\newcommand{\intall}{\int_{-\infty}^{\infty}}
\newcommand{\avg}[1]{\langle#1\rangle}
\newcommand{\Avg}[1]{\left\langle#1\right\rangle}

\newcommand{\bk}[1]{\left(#1\right)}
\newcommand{\Bk}[1]{\left[#1\right]}
\newcommand{\BK}[1]{\left\{#1\right\}}

\newcommand{\trace}{\operatorname{tr}}

\newcommand{\test}{\underset{\mathcal H_0}{\overset{\mathcal H_1}{\gtrless}}}

\begin{document}

\title{Volterra filters for quantum estimation and detection}

\author{Mankei Tsang}

\email{mankei@nus.edu.sg}
\affiliation{Department of Electrical and Computer Engineering,
  National University of Singapore, 4 Engineering Drive 3, Singapore
  117583}

\affiliation{Department of Physics, National University of Singapore,
  2 Science Drive 3, Singapore 117551}



\date{\today}


\begin{abstract}
  The implementation of optimal statistical inference protocols for
  high-dimensional quantum systems is often computationally
  expensive. To avoid the difficulties associated with optimal
  techniques, here I propose an alternative approach to quantum
  estimation and detection based on Volterra filters.  Volterra
  filters have a clear hierarchy of computational complexities and
  performances, depend only on finite-order correlation functions, and
  are applicable to systems with no simple Markovian model. These
  features make Volterra filters appealing alternatives to optimal
  nonlinear protocols for the inference and control of complex quantum
  systems. Applications of the first-order Volterra filter to
  continuous-time quantum filtering, the derivation of a
  Heisenberg-picture uncertainty relation, quantum state tomography,
  and qubit readout are discussed.
\end{abstract}

\maketitle

\section{Introduction}
The advance of quantum technologies relies on our ability to measure
and control complex quantum systems. An important task in quantum
control is to infer unknown variables from the noisy measurements of a
quantum system.  Examples include the prediction of quantum dynamics
for measurement-based feedback control
\cite{wiseman_milburn,gardiner_zoller,jacobs,bouten,*bouten09,haroche_raimond}
and the estimation and detection of weak signals
\cite{helstrom,holevo,braginsky,paris_rehacek,rbk10b,*ferrie14,riofrio,cook14,audenaert09,six,granade15,glm2011,smooth,*smooth_pra1,*smooth_pra2,*gammelmark2013,*guevara,hypothesis,testing_quantum,gambetta07,danjou,*danjou15,ng,wheatley,*yonezawa,*iwasawa}. To
implement the signal processing for such tasks, a Bayesian
decision-theoretic formulation of optimal quantum statistical
inference is now well established
\cite{helstrom,holevo,wiseman_milburn,jacobs,gardiner_zoller,bouten,*bouten09,haroche_raimond,
  smooth,*smooth_pra1,*smooth_pra2,*gammelmark2013,*guevara,hypothesis,testing_quantum}.
The quantum filtering theory pioneered by Belavkin \cite{belavkin89,[]
  [{, and references therein.}] belavkin} for the optimal prediction
of quantum dynamics has especially been hailed as a seminal
achievement in quantum control theory; its applications to
measurement-based cooling \cite{steck,*steck06}, squeezing
\cite{thomsen}, state preparation \cite{yanagisawa06,*negretti},
quantum error correction \cite{ahn,*sarovar,chase08}, qubit readout
\cite{gambetta07,danjou,*danjou15,ng}, and quantum state tomography
\cite{riofrio,cook14,audenaert09,six} in atomic, optical,
optomechanical, condensed-matter, and
superconducting-microwave-circuit systems \cite{wiseman_milburn} have
been studied extensively in the literature.

Although optimal quantum inference has been successful experimentally
for low-dimensional systems, such as qubits \cite{hume} and few-photon
systems \cite{sayrin}, as well as near-Gaussian systems, such as
optical phase estimation \cite{wheatley,*yonezawa,*iwasawa} and
optomechanics \cite{wieczorek}, its implementation for
high-dimensional non-Gaussian quantum systems is beset with
difficulties in practice. An exact implementation of the quantum Bayes
rule \cite{gardiner_zoller} for optimal inference requires numerical
updates of the posterior density matrix based on the measurement
record. Except for special cases such as Gaussian systems
\cite{wiseman_milburn}, the number of elements needed to keep track of
the density matrix scales exponentially with the degrees of freedom,
making the implementation prohibitive for many-body non-Gaussian
systems.  This problem, known as the curse of dimensionality, means
that approximations must often be sought
\cite{steck,*steck06,vladimirov,ahn,*sarovar,chase08,amini,hush,nielsen09}. Current
approximation techniques for dynamical systems include Gaussian
approximations \cite{steck,*steck06,audenaert09,vladimirov},
phase-space particle filters \cite{hush}, Hilbert-space truncation
\cite{chase08,amini}, and manifold learning \cite{nielsen09}, but
these techniques provide little assurance about their actual errors
and often remain too expensive to compute for real-time control of
high-dimensional systems. Another problem with optimal inference and
the associated stochastic-master-equation approach is its reliance on
a Markovian model, which is difficult to use for many complex systems,
especially those with $1/f$ or fractional noise statistics.  With the
ongoing trend of increasing complexity in quantum experiments, not
only with condensed matter but also with optomechanics
\cite{aspelmeyer14}, atomic ensembles \cite{bloch}, and
superconducting circuits \cite{houck}, optimal inference is becoming
an unattainable goal in practice.

Against this backdrop, here I propose an alternative approach to
quantum estimation and detection based on Volterra filters.  Instead
of seeking absolute optimality, Volterra filters are a class of
polynomial estimators with a clear hierarchy of computational
complexities and estimation errors \cite{mathews_sicuranza}. Their
applications to quantum estimation and detection promise to solve many
of the practical problems associated with optimal quantum inference,
including the curse of dimensionality, the lack of error assurances
upon approximations, and the need for a Markovian model. The filter
errors also provide a set of upper error bounds on the Bayesian
quantum Cram\'er-Rao \cite{helstrom,holevo,twc}, Ziv-Zakai
\cite{qzzb,*qbzzb}, and Helstrom
\cite{helstrom,holevo,tsang_nair,*tsang_open} bounds, forming novel
hierarchies of fundamental uncertainty relations and may be of
independent foundational interest. The Volterra series has recently
been used to model the input-output relations of a quantum system
\cite{zhang_volterra14}, but my focus here is different and concerns
the estimation of hidden observables and hypothesis testing given the
output measurement record.

\section{Quantum estimation}
\subsection{Formalism}
Consider a quantum system in the Heisenberg picture with initial
density operator $\rho$.  Let
\begin{align}
y = \bk{\begin{array}{c}y(1)\\ y(2)\\ \vdots \\ y(K)\end{array}}
\end{align}
be a column vector of observables under measurement.  For example, $y$
can be the observables of an output optical field under homodyne,
heterodyne, or photon-counting measurements.  Given a measurement
record of $y$, the goal of quantum estimation is to infer a column
vector of hidden observables 
\begin{align}
x \equiv
\bk{\begin{array}{c}x(1)\\ x(2)\\ \vdots\\ x(J)\end{array}}.
\end{align}
For example, $x$ can be the observables of a quantum system that has
interacted with the optical field, such as the position of a quantum
mechanical oscillator or a spin operator of an atomic ensemble, and
the goal of the estimation is to infer $x$ given the measurement
record.  Quantum estimation is usually framed in the Schr\"odinger
picture via the concept of posterior density operator
\cite{wiseman_milburn,gardiner_zoller}, but it can be shown to be
equivalent to the Heisenberg-picture approach adopted here
\cite{bouten,*bouten09}. This task is especially important for
measurement-based feedback control \cite{wiseman_milburn}, such as
measurement-based cooling and squeezing, to gain real-time information
about quantum degrees of freedom and to reduce their uncertainties via
feedback control. Experiments that implement quantum estimation have
been reported in Refs.~\cite{hume,sayrin,wieczorek} for example.

The estimation error has a well-defined decision-theoretic meaning if
all the $x$ and $y$ operators commute with one another, such that $x$
and $y$ can be jointly measured and treated as classical random
variables in the same probability space
\cite{holevo,bouten,*bouten09,belavkin_qnd}.  This assumption is applicable to a
wide range of scenarios, including quantum filtering
\cite{bouten,*bouten09,belavkin_qnd} and the estimation of any classical
parameter or waveform coupled to a quantum system
\cite{smooth,*smooth_pra1,*smooth_pra2,*gammelmark2013,*guevara,gao}.

Since $x$ and $y$ are compatible observables, the rest of the
estimation theory is identical to the classical treatment
\cite{mathews_sicuranza}. Let $\check x(j|y)$ be an estimator of
$x(j)$ given $y$, and assume that the estimator is given by the
truncated Volterra series, viz.,
\begin{align}
\check x(j|y) &= \sum_{p=0}^P \sum_{1\le k_{1} \le k_2 \le \dots \le k_{p} \le K} 
h_p(j,k_{1},k_{2},\dots,k_{p}|\theta)
\nonumber\\&\quad\times y(k_{1})y(k_{2})\dots y(k_{p}),
\label{volterra}
\end{align}
where $\theta$ is a vector of tunable parameters, $P$ is the order of
the series and quantifies the complexity of the filter, and the
zeroth-order term is simply a constant $h_0(j)$ and does not depend on
$y$. For $P\to\infty$, the series can be regarded as the Taylor series
for an arbitrary estimator, although I will focus on finite $P$.

A useful trick to simplify the notations is to define the set of all
products of $y$ elements up to order $P$ as
\begin{align}
y^{(P)} \equiv \BK{1,y,y^{\otimes 2},\dots, y^{\otimes P}},
\end{align}
where
\begin{align}
y^{\otimes p} &\equiv \left\{y(k_1)y(k_2)\dots y(k_p); 
\right.
\nonumber\\&\quad
\left.
1\le k_1 \le k_2\le \dots \le k_p \le K\right\}
\end{align}
is the set of all $p$th-order products of $y$ elements. Then the
Volterra series in Eq.~(\ref{volterra}) can be rewritten as
\begin{align}
\check x(j|y) &= \sum_\mu h^{(P)}(\mu|\theta) y^{(P)}(\mu),
\end{align}
where $h^{(P)}$ is a linear filter with respect to $y^{(P)}$ but
equivalent to the Volterra filter that is nonlinear with respect to
$y$, and $\mu$ is a composite index that goes through all elements in
$y^{(P)}$.

Define 
\begin{align}
\Avg{f(x,y)} \equiv \trace\Bk{\rho f(x,y)}
\end{align}
as the expectation of any function of $x$ and $y$, with $\trace$
denoting the operator trace. Let the error covariance matrix be
\begin{align}
\Sigma(j,k) \equiv \Avg{\Bk{x(j)-\check x(j|y)}\Bk{x(k)-\check x(k|y)}}.
\end{align}
The absolutely minimum mean-square error for arbitrary estimators is
achieved by the conditional expectation of $x$ given $y$
\cite{bouten,*bouten09}. For the optimal filtering and prediction of quantum
observables for example, the usual method is to compute the posterior
density operator $\rho(y)$ conditioned on the measurement record $y$
in the Schr\"odinger picture using the Kraus operators that
characterize the measurements \cite{wiseman_milburn,gardiner_zoller},
and then take the conditional expectation given by
$\check x(j|y) = \trace[x_S(j) \rho(y)]$, with $x_S(j)$ being the
Schr\"odinger picture of $x(j)$. If the continuous-time limit is
taken, the posterior density operator obeys the celebrated stochastic
master equation \cite{wiseman_milburn,gardiner_zoller,jacobs,bouten,*bouten09}
first proposed by Belavkin \cite{belavkin89,belavkin}.  The
computation of $\rho(y)$ suffers from the curse of dimensionality
however. To restrict the complexity, consider here instead the error
of the $P$th-order Volterra filter given by
\begin{align}
\Sigma^{(P)}(j,k|\theta) &=
\left\langle\Bk{x(j)-\sum_\mu h^{(P)}(j,\mu|\theta) y^{(P)}(\mu)}
\right.
\nonumber\\&\quad\times
\left.
\Bk{x(k)-\sum_\mu h^{(P)}(k,\mu|\theta) y^{(P)}(\mu)}
\right\rangle
\\
&= C_x(j,k)-\sum_\mu h^{(P)}(j,\mu|\theta) C_{xy^{(P)}}(k,\mu)
\nonumber\\&\quad
-\sum_\mu h^{(P)}(k,\mu|\theta) C_{xy^{(P)}}(j,\mu)
\nonumber\\&\quad
+\sum_{\mu,\nu}h^{(P)}(j,\mu|\theta)h^{(P)}(k,\nu|\theta) C_{y^{(P)}}(\mu,\nu),
\label{mse_volterra}
\end{align}
where 
\begin{align}
C_x(j,k) &\equiv \Avg{x(j)x(k)},
\\
C_{xy^{(P)}}(j,\mu) &\equiv \Avg{x(j) y^{(P)}(\mu)},
\\
C_{y^{(P)}}(\mu,\nu) &\equiv \Avg{y^{(P)}(\mu)y^{(P)}(\nu)}.
\end{align}
To optimize the Volterra filter, one can seek the parameters $\theta$
that minimize any desired component of $\Sigma^{(P)}(j,k|\theta)$ in
Eq.~(\ref{mse_volterra}), which has the remarkable feature of
depending only on finite-order correlations.  Specifically,
$C_{xy^{(P)}}(j,\mu)$ depends on the correlation between $x(j)$ and
products of $y$ elements up to the $P$th order, and $C_{y^{(P)}}$
depends on the correlations among $y$ up to the $2P$th
order. Stationarity assumptions and frequency-domain techniques can
further simplify the expressions.

Quantum mechanics comes into the problem through the correlations.
They must obey uncertainty relations with other incompatible
observables \cite{holevo,ozawa03}. They can violate Bell
\cite{speakable,*horodecki} and Leggett-Garg
\cite{leggett_garg,*emary2014} inequalities, requiring different
probability spaces for different experimental settings.  They may
result from nontrivial internal quantum dynamics with no classical
correspondence; the promise of quantum computation and simulation
\cite{nielsen} is in fact based on the difficulty of reproducing
quantum dynamical statistics using any hidden-variable model. This
difficulty also means that attempts to simplify quantum filters via
classical models \cite{steck,*steck06,vladimirov,hush} are likely to be
inaccurate for highly nonclassical systems. The Volterra filters
sidestep the issue via a manifestly non-Markovian approach that does
not require an \emph{online} simulation of the internal quantum
dynamics.  The identification of the correlations and the filter
synthesis, though nontrivial, can be done \emph{offline} for control
applications.

A challenge for classical applications of Volterra filters is that the
correlations are often difficult to model or measure in practice, but
it is less problematic for quantum systems: computing and measuring
correlation functions is already a major endeavor in condensed-matter
physics \cite{weiss,*datta,*bruus} and early quantum optics
\cite{mandel} with an extensive literature.  The Volterra-series
approach to input-output analysis \cite{zhang_volterra14} should also
help their simulation. Compared with the stochastic-master-equation
approach
\cite{wiseman_milburn,gardiner_zoller,jacobs,bouten,*bouten09}, the
use of correlation functions has the advantage of not requiring a
Markovian model or stochastic calculus, although the Volterra filters
may require a longer memory depending on the time scales of the
correlation functions and the signal-to-noise properties. An empirical
alternative to prior system identification is to train the filter
directly using experimental or simulated data to minimize the sample
errors.

I now consider the ideal case where arbitrary Volterra filters can be
implemented, such that the tunable parameters $\theta$ are all
elements of $h^{(P)}$.  Since $\Sigma^{(P)}$ is quadratic with respect
to $h^{(P)}$, the minimization can be performed analytically.  Define
the risk function \cite{berger} to be minimized as
\begin{align}
R(\theta) &\equiv \sum_{j,k} u(j)\Sigma^{(P)}(j,k|\theta) u(k),
\end{align}
where $u$ is an arbitrary real vector. The
optimal Volterra filter
\begin{align}
\tilde h^{(P)} \equiv \arg \min_{h^{(P)}} R(h^{(P)})
\end{align}
for arbitary $u$ satisfies the equation
\begin{align}
C_{xy^{(P)}}(j,\nu) &= \sum_\mu \tilde h^{(P)}(j,\mu)C_{y^{(P)}}(\mu,\nu),
\label{wiener_hopf}
\end{align}
which is a system of linear equations with respect to $\tilde h^{(P)}$
and can be solved by conventional methods, and the resulting error
covariance matrix is
\begin{align}
\tilde\Sigma^{(P)}(j,k)  &\equiv \Sigma^{(P)}(j,k|\tilde h^{(P)})
\\
&= C_x(j,k) - \sum_\mu \tilde h^{(P)}(j,\mu)C_{xy^{(P)}}(k,\mu).
\label{mmse_volterra}
\end{align}
This error can be computed offline to evaluate the optimal performance
of a Volterra filter and the trade-off between the error and the
filter complexity $P$. Going to a higher order is guaranteed not to
increase the error, since $\tilde\Sigma^{(P)} \le \tilde\Sigma^{(Q)}$
if $P > Q$ (a higher-order filter can always achieve the performance
of a lower-order filter by ignoring the higher-order terms in
$y^{(P)}$). As the infinite-order Volterra filter can be regarded as
the Taylor series for an arbitrary function, $\tilde h^{(\infty)}$
will be the optimal among arbitrary estimators and
$\tilde\Sigma^{(\infty)}$ will coincide with the absolutely optimal
error. $\tilde\Sigma^{(P)}$ thus provides a hierarchy of increasingly
tight upper error bounds for optimal quantum inference. 
Most importantly, a finite-order Volterra filter can still enjoy a
performance given by Eq.~(\ref{mmse_volterra}) for any statistics,
even if it is not optimal in the absolute sense. On a fundamental
level, it is interesting to note that, if $x$ is classical, the upper
error bounds also apply to the Bayesian quantum Cram\'er-Rao
\cite{helstrom,holevo,twc} and Ziv-Zakai \cite{qzzb,*qbzzb} lower
error bounds, forming a novel set of operationally motivated
uncertainty relations; an example is shown in Sec.~\ref{relation}.

The optimal $P = 0$ Volterra filter does not process the measurement
and is simply given by the prior expectation $\avg{x}$.  The $P = 1$
Volterra filter is a linear filter with respect to $y$ and deserves
special attention, as it is the simplest Volterra filter beyond the
trivial zeroth-order case and will likely become the most popular. If
$x$ and $y$ are jointly Gaussian, the optimal linear filter is also
the optimal among arbitrary estimators and equivalent to the Kalman
filter when applied to the prediction of Markovian dynamical systems
\cite{vantrees}, but the linear filter can still be used for any
non-Gaussian or non-Markovian statistics and depends only on the
second-order correlations in terms of $x$ and $y$.

\subsection{Continuous-time quantum filtering}
For example, consider the continuous-time quantum filtering and
prediction problem, which is to estimate a Heisenberg-picture
observable $x(t)$ given the past measurement record
$\{y(\tau); t_0\le \tau \le T < t\}$ \cite{bouten,*bouten09}. It can be shown
that all the Heisenberg-picture operators under consideration commute
with one another under rather general conditions for filtering and
prediction \cite{bouten,*bouten09,belavkin_qnd}. If $t < T$ is desired for
smoothing
\cite{smooth,*smooth_pra1,*smooth_pra2,*gammelmark2013,*guevara}, care
should be taken in the modeling to ensure that $x(t)$ still commutes
with $y$ and an operational meaning of the estimation error
exists. For example, a c-number signal, such as a classical force,
commutes with all operators by definition.

To transition from the discrete formalism to continous time,
define a discrete time given by
\begin{align}
  t_j &= t_0 + j\delta t,
\label{discrete_time}
\end{align}
with initial time $t_0$, integer $j$, and time
interval $\delta t$.  For infinitesimal $\delta t$, the linear $P = 1$
estimator in the continuous-time limit becomes
\begin{align}
\check x(t|y) &= h_0(t) + \int_{t_0}^T d\tau h_1(t,\tau) y(\tau),
\label{linear_filter}
\end{align}
where $\check x(t|y)$, $h_0(t)$, $h_1(t,\tau)$, and $y(\tau)$ are
continuous-time versions of $\check x(j|y)$, $h_0(j)$,
$h_1(j,k)/\delta t$, and $y(k)$, respectively.
Eq.~(\ref{linear_filter}) is a continuous-time limit of the Volterra
series in Eq.~(\ref{volterra}) for $P = 1$.  Assuming zero-mean $x$
and $y$ for simplicity and using Eqs.~(\ref{wiener_hopf}) and
(\ref{mmse_volterra}), the optimal linear filter
$\tilde h_{1}(t,\tau)$ and the corresponding mean-square error
$\tilde\Sigma^{(1)}(t,t)$ can be expressed as
\begin{align}
C_{xy}(t,\tau) &= \int_{t_0}^T ds \tilde h_{1}(t,s) C_y(s,\tau),
  \\
\tilde\Sigma^{(1)}(t,t) &= C_x(t,t) - 
\int_{t_0}^T d\tau \tilde h_{1}(t,\tau) C_{xy}(t,\tau),
\end{align} 
where 
\begin{align}
C_x(t,t) &\equiv \Avg{x^2(t)},
\\
C_{xy}(t,\tau) &\equiv \Avg{x(t)y(\tau)},
\\
C_y(t,\tau) &\equiv \Avg{y(t)y(\tau)}
\end{align}
are the only correlation functions needed to compute both the filter
and the error. Although this form of the optimal linear estimator is
known in the classical context \cite{vantrees}, its applicability to
quantum systems with any nonlinear dynamics and non-Gaussian
statistics is hitherto unappreciated. Compared with the stochastic
master equation, the linear filter can be more easily implemented
using fast digital electronics or even analog electronics in practice
\cite{wheatley,*yonezawa,*iwasawa,stockton02} for measurement-based
feedback control, while the implementation of higher-order filters is
more involved but can leverage existing digital-signal-processing
techniques \cite{mathews_sicuranza}.

\subsection{\label{relation}Heisenberg-picture uncertainty relation}
To demonstrate a side consequence of the Volterra-filter formalism,
here I use the analytic error expression for the first-order Volterra
filter to derive a quantum uncertainty relation for Heisenberg-picture
operators. Consider the Hamiltonian
$\mathfrak H(t) = \mathfrak H_0(t) - q x(t)$, where $q$ is a canonical
position operator, $x(t)$ is a classical force, and $\mathfrak H_0$ is
the rest of the Hamiltonian. Suppose that $\mathfrak H_0$ is at most
quadratic with respect to canonical position and momentum operators,
such that the equations of motion for those operators in the
Heisenberg picture are linear. The initial density operator $\rho$, on
the other hand, can have any non-Gaussian statistics.

Consider an output field quadrature operator $y(t)$ that commutes with
itself at different times in the Heisenberg picture
\cite{bouten,*bouten09}. For example, it can model the homodyne
measurement of an output optical field in optomechanics.  It can be
shown that
\begin{align}
y(t) &= y_0(t) + \int_0^T dt g(t,\tau)x(\tau),
\label{green}
\end{align}
where 
\begin{align}
g(t,\tau) &= \left\{
\begin{array}{ll}
\frac{i}{\hbar}\Bk{y_0(t),q_0(\tau)}, & t > \tau,
\\
0, & t \le \tau,
\end{array}
\right.
\end{align}
is the causal c-number commutator and the subscript $0$ denotes the
interaction picture with respect to the Hamiltonian $\mathfrak H_0$.

Without loss of generality, assume that $x(t)$, $y_0(t)$, and $q_0(t)$
are zero-mean processes. Consider the estimation of $x(t)$ using the
record $\{y(\tau); 0 < \tau \le T\}$.  If $y_0(t)$ has non-Gaussian
statistics, the optimal nonlinear estimator is difficult to derive,
but the first-order Volterra filter given by
\begin{align}
\check x(t|y) &= \int_0^T d\tau h_1(t,\tau)y(\tau)
\label{estimator}
\end{align}
can be analyzed more easily.  To proceed, it is more convenient to
consider discrete time as defined in
Eq.~(\ref{discrete_time}). Regarding $x$, $y_0$, $y$, and $\check x$
as column vectors and $g$ and $h_1$ as matrices, Eqs.~(\ref{green})
and (\ref{estimator}) can be rewritten in matrix form as
\begin{align}
y &= y_0 + \delta t g x,
\\
\check x &= \delta t h_1 y.
\end{align}
With covariance matrices defined as
\begin{align}
C_x &\equiv \Avg{xx^\top},
\\
C_{y0} &\equiv \Avg{y_0y_0^\top},
\\
C_{xy} &\equiv \Avg{xy^\top} = \delta t C_x g^\top, 
\\
C_y &\equiv \Avg{yy^\top} = \delta t^2 g C_x g^\top + C_{y0},
\end{align}
where $\top$ denotes the matrix transpose, the optimal linear filter becomes
\begin{align}
\delta t\tilde h_1  &= C_{xy} C_y^{-1} = 
\delta t  C_x g^\top\bk{\delta t^2 g C_x g^\top + C_{y0}}^{-1},
\end{align}
and the error covariance matrix becomes
\begin{align}
\tilde\Sigma^{(1)} &\equiv \Avg{\bk{x-\delta t\tilde h_1 y}\bk{x-\delta t\tilde h_1 y}^\top}
\\
&= C_x - \delta t \tilde h_1 C_{xy}^\top
\\
&= C_x - \delta t^2  C_x g^\top\bk{\delta t^2 gC_x g^\top + C_{y0}}^{-1} g C_x
\\
&= \bk{C_x^{-1} + \delta t^2 g^\top C_{y0}^{-1} g}^{-1},
\label{mse}
\end{align}
where the last line uses the matrix inversion lemma \cite{simon}.

The error covariance can be compared with the Bayesian quantum
Cram\'er-Rao bound derived in Ref.~\cite{twc}. The quantum bound for
Gaussian $x$ results in a matrix inequality given by
\begin{align}
\tilde\Sigma^{(1)} &\ge 
\bk{C_x^{-1} + \frac{4\delta t^2}{\hbar^2} C_{q0}}^{-1},
\label{qcrb}
\end{align}
where 
\begin{align}
C_{q0}(t_j,t_k) &\equiv 
\frac{1}{2}\Avg{q_0(t_j)q_0(t_k) + q_0(t_k)q_0(t_j)}.
\end{align}
Unlike $x(t)$ and $y(t)$, $q(t)$ may not self-commute at different
times, and the symmetric ordering in the covariance function
\cite{clerk} arises naturally from the derivation of the quantum bound
in Ref.~\cite{twc}. Comparing Eq.~(\ref{mse}) and Eq.~(\ref{qcrb}), it
can be seen that the inequality holds only if
\begin{align}
g^\top C_{y0}^{-1}g &\le \frac{4}{\hbar^2} C_{q0},
\label{matrix_qcrb}
\end{align}
which is a matrix uncertainty relation between two quantum processes
in the Heisenberg picture involving their causal commutator $g$.  Note
that $y_0$ and $q$ are canonical phase-space coordinate operators with
linear dynamics but need not have Gaussian statistics. The end result
does not involve $x$ and can be applied to any quantum system that
satisfies the stated assumptions beyond the estimation scenario. The
estimation procedure nonetheless gives the relation a clear
operational meaning.

Eq.~(\ref{matrix_qcrb}) can be further simplified by assuming
linear-time-invariant dynamics and stationary statistics.  The result
in the continuous long-time limit is a spectral uncertainty relation
given by
\begin{align}
S_{y0}(\omega) S_{q0}(\omega) &\ge \frac{\hbar^2}{4} |G(\omega)|^2,
\end{align}
with the frequency-domain quantities defined by
\begin{align}
C_{y0}(t,\tau) &= \intall
\frac{d\omega}{2\pi} S_{y0}(\omega) \exp\Bk{i\omega(t-\tau)},
\\
C_{q0}(t,\tau) &= \intall
\frac{d\omega}{2\pi} S_{q0}(\omega) \exp\Bk{i\omega(t-\tau)},
\\
g(t,\tau) &= \intall
\frac{d\omega}{2\pi} G(\omega) \exp\Bk{i\omega(t-\tau)}.
\end{align}
The spectral relation imposes a lower bound on the noise floor of an
output operator $y_0(t)$ in terms of the spectrum of a noncommuting
operator $q_0(t)$. For example, the relation can be used to determine
the fundamental limit to the noise floor of optical homodyne detection
as a function of the mechanical-position power spectral density for a
gravitational-wave detector \cite{braginsky,miao15}. The inequality
can be saturated if the quantum statistics are Gaussian \cite{twc}.

\subsection{Quantum state tomography}
For an application in quantum information processing, consider
the estimation of parameters in a density matrix, also known as
quantum state tomography
\cite{paris_rehacek,rbk10b,*ferrie14,riofrio,cook14,audenaert09,six,granade15}. Assume
a $d\times d$ density matrix of the form
\begin{align}
\rho_z &= \frac{I}{d} + \sum_{\alpha = 1}^{d^2-1} z_\alpha E_\alpha,
\end{align}
where $I$ is the identity matrix, $E_{\alpha}$ is a set of Hermitian,
traceless, and orthonormal matrices that satisfy
\begin{align}
E_\alpha &= E_\alpha^\dagger, &\trace E_\alpha &= 0,
&
\trace E_\alpha E_\beta &= \delta_{\alpha\beta},
\end{align}
and $z$ is a column vector of real unknown parameters.  $\rho_z$ is
Hermitian and $\trace \rho_z = 1$ by construction, and the density
matrix describes a physical quantum state only if $\rho_z$ is
positive-semidefinite \cite{riofrio}.  Measurements can often be
modeled as \cite{riofrio}
\begin{align}
y &= A z + y_0,
\end{align}
where $y$ is a column vector, $A$ is a known measurement matrix, and
$y_0$ is a zero-mean noise vector. The main difficulty with the
Bayesian estimation protocol
\cite{audenaert09,rbk10b,*ferrie14,granade15} is that, owing to the
physical-state requirement, the prior for $z$ is highly non-Gaussian,
while the statistics of $y_0$ may also be non-Gaussian. With the
non-Gaussian statistics and $d$ scaling exponentially with the degrees
of freedom, exact Bayesian estimation of $z$ would suffer from the
curse of dimensionality.  Existing approximation techniques include
Gaussian approximations \cite{audenaert09} and particle filters
\cite{granade15}, but their actual estimation errors remain unclear.

The Volterra filters can be used despite the non-Gaussianity of $z$ or
$y_0$. Let
\begin{align}
x &= B z
\end{align}
be a column vector of parameters to be estimated for a given sampling
matrix $B$. Note that $B$ can be a non-square matrix and the number of
elements in $x$ can be much smaller than that in $z$ if the
dimensionality of the latter is a concern. For example, the fidelity
between the density matrix and a target pure state \cite{flammia} can
be expressed in this way, in which case $B$ is a row vector and $x$ is
a scalar.  The optimal first-order filter can be expressed as
\begin{align}
\check x &= B\Avg{z} + \tilde h_1 \bk{y- A\Avg{z}},
\label{qst_filter}
\\
  \tilde h_1 &= B C_z A^\top \bk{A C_z A^\top + C_{y0}}^{-1},
  \\
  C_z &\equiv \Avg{z z^\top}- \Avg{z}\Avg{z}^\top.
\end{align}
The filter is guaranteed to offer an error
covariance matrix given by
\begin{align}
\tilde\Sigma^{(1)} &= B\bk{C_z^{-1} + A^\top C_{y0}^{-1} A}^{-1} B^\top.
\end{align}
The linear complexity and the error guarantee are the main advantages
of the Volterra filter. A shortcoming is that, due to noise and the
lack of a constraint in the algorithm, the estimate $\check x$ may not
lead to a positive-semidefinite density matrix. If this is a problem,
an obvious remedy is to find the physical $x$ closest to $\check x$
with respect to a distance measure. A more sophisticated way is to
compute the posterior distribution over a region near $\check x$ with
a volume suggested by $\tilde\Sigma^{(1)}$. If the noise is low enough
or the number of trials is large enough such that $\tilde\Sigma^{(1)}$
is small, the region needs to cover a small parameter subspace only,
and the curse of dimensionality can be avoided.

The remaining issue is the choice of prior $\avg{z}$ and $C_z$ in an
objective manner. One option is to take one of the commonly used
objective priors for $z$ \cite{riofrio,granade15} and compute its
moments. For $d = 2$ and $z$ being the Bloch vector, the prior moments
can be easily calculated by taking advantage of the Bloch spherical
symmetry. The computation seems nontrivial for $d \ge 3$, but for each
$d$ it needs to be done just once and for all.

The most conservative and arguably paranoid option is to choose a
prior that is least favorable to the Volterra filter. Given a prior
probability measure $\pi_z$ on $z$, one can define a risk function,
such as the Hilbert-Schmidt distance given by
\begin{align}
R(\pi_z) &= \trace \tilde\Sigma^{(1)}(\pi_z).
\end{align}
Then the least favorable prior is one that maximizes the risk while
still observing the physical constraint on $\rho_x$, that is,
\begin{align}
\arg \max_{\pi_z; \rho_z \ge 0} R(\pi_z).
\end{align}
Note that this prior depends in general on the measurement matrix $A$
as well as the sampling matrix $B$. Without the physical constraint,
the least favorable $C_z$ would be infinite, giving
\begin{align}
\tilde\Sigma^{(1)} &\le B\bk{A^\top C_{y0}^{-1} A}^{-1} B^\top,
\label{minimax_bound}
\end{align}
and the Volterra filter would become equivalent to the unconstrained
maximum-likelihood estimator for Gaussian $y_0$. The effect of a
finite $C_z$ is to pull the estimate from the maximum-likelihood value
towards the prior $\avg{x}$ via the weighted average given by
Eq.~(\ref{qst_filter}).

\section{Quantum detection}
\subsection{Formalism}
Assume two hypotheses denoted by $\mathcal H_0$ and $\mathcal H_1$.
These hypotheses can be about the initial density operator as well as
the dynamics and measurements of the quantum system
\cite{hypothesis}. As before, let the measured Heisenberg-picture
observables be $y$ with commuting elements under both hypotheses. The
goal of detection is equivalent to binary hypothesis testing, which is
to make a decision on $\mathcal H_0$ or $\mathcal H_1$ based on
$y$. Applications include force detection
\cite{braginsky,hypothesis,tsang_nair,*tsang_open}, fundamenal tests
of quantum mechanics
\cite{hypothesis,testing_quantum,chen2013,aspelmeyer14}, quantum error
correction \cite{ahn,*sarovar,wiseman_milburn}, and qubit readout
\cite{gambetta07,danjou,*danjou15,ng}.  Prior work on the use of
Volterra filters for classical detection focuses on the heuristic
deflection criterion \cite{picinbono90,picinbono95}, but it does not
seem to have any decision-theoretic meaning or relationship with the
more rigorous criteria of error probabilities \cite{picinbono95}. Here
I propose a similar performance criterion that is able to provide an
upper bound on the average error probability, while still offering a
simple design rule for the Volterra filters. To my knowledge the
proposed design rule is new also in the context of classical detection
theory.

Let $\lambda(y)$ be a test statistic as a polynomial function of $y$
similar to Eq.~(\ref{volterra}). For later notational convenience, I
will rewrite it as
\begin{align}
\lambda(y) &= h_0 + H^\top Y,
\end{align}
where the zeroth-order term $h_0$ is written separately, $Y$ is a
column vector with the elements in
$\{y,y^{\otimes 2},\dots, y^{\otimes P}\}$ without the constant term
$1$, $H$ is a column vector with the corresponding elements in
$h^{(P)}$, and $\top$ denotes the transpose.  Let $\avg{f(y)}_0$ be the
expectation of a function of $y$ given hypothesis $\mathcal H_0$, and
$\avg{f(y)}_1$ be the expectation given hypothesis $\mathcal H_1$.
Note that the hypotheses can be about the initial density operator,
the dynamics, and the definition of $y$.

I demand the test statistic to have different expectations for the two
hypotheses, viz., 
\begin{align}
\Avg{\lambda}_0 \neq \Avg{\lambda}_1.
\end{align}
This means
that the order $P$ cannot be arbitrary but must be high enough to
result in different expectations. I further demand the expectations to
be symmetric around $0$, viz.,
\begin{align}
\Avg{\lambda}_0 + \Avg{\lambda}_1 = 0.
\end{align}
This is accomplished by setting
\begin{align}
h_0 &= - H^\top \bar Y,
\\
\bar Y &\equiv \frac{1}{2}\bk{\Avg{Y}_0+\Avg{Y}_1},
\end{align}
resulting in 
\begin{align}
\Avg{\lambda}_1 &=  -\Avg{\lambda}_0 = H^\top \Delta,
\\
\Delta &\equiv \frac{1}{2} \bk{\Avg{Y}_1-\Avg{Y}_0}.
\end{align}
Without loss of generality, I assume
$\avg{\lambda}_1 = H^\top\Delta > 0$. Consider a threshold test that
decides on $\mathcal H_0$ if $\lambda < 0$ and $\mathcal H_1$ if
$\lambda \ge 0$.  This is commonly expressed as \cite{vantrees}
\begin{align}
\lambda(y) \test 0.
\end{align}
The average error probability becomes
\begin{align} \mathcal P_e(H) &= \pi_0\Avg{1_{\lambda \ge 0}(y)}_0 +  
\pi_1\Avg{1_{\lambda < 0}(y)}_1,
\end{align}
where $\pi_0$ and $\pi_1$ are the prior probabilities for the
hypotheses and $1_{\lambda \ge 0}$ and $1_{\lambda < 0}$ are indicator
functions.  Since $\mathcal P_e(H)$ in general depends on infinite
orders of $\lambda$ moments, I appeal to the Cantelli inequality
\cite{billingsley} to obtain
\begin{align}
\Avg{1_{\lambda \ge 0}(y)}_0 \le
\frac{\Avg{\lambda^2}_0-\Avg{\lambda}_0^2}{\Avg{\lambda^2}_0},
\end{align}
and similarly for $\avg{1_{\lambda < 0}(y)}_1$. This leads to upper
bounds on $\mathcal P_e$ given by
\begin{align}
  \mathcal P_e(H) &\le \mathcal Q(H) \le \mathcal R(H),
\label{Pebounds}\\
\mathcal Q(H) &\equiv \frac{\pi_0}{1+(H^\top \Delta)^2/(H^\top C_0H)}
\nonumber\\&\quad
+\frac{\pi_1}{1+(H^\top \Delta)^2/(H^\top C_1H)},
\label{Q}
\\
\mathcal R(H) &\equiv 
\frac{H^\top (\pi_0 C_0 + \pi_1 C_1)H}{(H^\top \Delta)^2},
\label{R}
\end{align}
where
\begin{align}
C_{0} &\equiv \Avg{YY^\top}_0 - \Avg{Y}_0\Avg{Y}_0^\top,
\\
C_1 &\equiv \Avg{YY^\top}_1 - \Avg{Y}_1\Avg{Y}_1^\top
\end{align}
are the conditional covariance matrices.  $1/\mathcal R$ can be
regarded an output signal-to-noise ratio and has a similar form to the
deflection criterion \cite{picinbono90,picinbono95}, although
$\mathcal R$ has a clearer decision-theoretic meaning as an upper
error bound.

The purpose of using $\mathcal R$ rather than $\mathcal P_e$ or
$\mathcal Q$ is to define an easy-to-optimize criterion in terms of
finite-order correlations. To find the $\mathcal R$-optimal filter,
consider the Cauchy-Schwarz inequality
\begin{align}
\bk{H^\top\Delta}^2 \le \bk{H^\top M H}\bk{\Delta^\top M^{-1} \Delta}
\end{align}
for any positive-definite matrix $M$. The inequality is saturated if
and only if $H = \alpha M^{-1}\Delta$ for any constant
$\alpha$. Setting $M = \pi_0 C_0 + \pi_1 C_1$, I obtain
\begin{align}
  \tilde R&\equiv \min_{H}\mathcal R(H) 
= \frac{1}{\Delta^\top (\pi_0 C_0 + \pi_1 C_1)^{-1}\Delta},
  \\
  \tilde H &\equiv \arg \min_H \mathcal R(H) 
             =\alpha(\pi_0 C_0 + \pi_1 C_1)^{-1}\Delta,
\end{align}
and the $\mathcal R$-optimal test statistic
$\tilde\lambda(y)\equiv h_0+\tilde H^\top Y$, taking $\alpha = 1$
without loss of generality, becomes
\begin{align}
\tilde\lambda(y) 
=\Delta^\top(\pi_0 C_0 + \pi_1 C_1)^{-1}  \bk{Y-\bar Y},
\end{align}
which can then be used in a threshold test. The merits of this
approach are similar to those in the estimation scenario: dependence
of $\tilde\lambda(y)$ on finite-order correlations $\Delta$, $C_0$,
and $C_1$ without relying on a Markovian model, a performance
guaranteed by upper bounds
$\mathcal P_e(\tilde H) \le \mathcal Q(\tilde H) \le \tilde{\mathcal
  R}$
(the actual $\mathcal P_e$ may be much lower), and a hierarchy of
decreasing $\tilde R$ versus increasing complexity.  For the study of
fundamental quantum metrology, $\mathcal P_e(\tilde H)$,
$\mathcal Q(\tilde H)$, and $\tilde{\mathcal R}$ also provide a set of
upper bounds on the Helstrom bound
\cite{helstrom,holevo,tsang_nair,*tsang_open}.

It is not difficult to show that, if the hypotheses are about the mean
of a Gaussian $y$ and $C_0 = C_1$, $\tilde\lambda(y)$ for $P = 1$
coincides with the well known matched filter, and the threshold test
of $\tilde\lambda(y)$ against $0$ leads to the optimal $\mathcal P_e$
among all decision rules if $\pi_0 = \pi_1$ \cite{vantrees}.  The
derivation of the $\mathcal R$-optimal Volterra filter here in fact
resembles the historic derivation of the linear matched filter via
maximizing an output signal-to-noise ratio \cite{vantrees}. The
crucial differences are that here $\tilde\lambda(y)$ can include
higher-order products of $y$ elements and the upper error bounds
provide performance guarantees even for non-Gaussian statistics.

\subsection{Qubit readout}
For an application of the detection theory, consider the qubit
readout problem described in
Refs.~\cite{gambetta07,danjou,*danjou15,ng}.  The goal is to infer the
initial state of the qubit in one of the two possibilities from noisy
measurements. The two hypotheses can be modeled as
\begin{align}
\mathcal H_0: y(t_k) &= y_0(t_k),
\nonumber\\
\mathcal H_1: y(t_k) &= S x(t_k) + y_0(t_k),
\label{hypotheses}
\end{align}
where $x$ is a hidden qubit observable that can undergo spontaneous
decay or excitation in time, $S$ is a positive signal amplitude, and
$y_0$ is a zero-mean noise process. To perform hypothesis testing
given a record of $y$, consider the first-order $\mathcal R$-optimal
decision rule given by
\begin{align}
\tilde H &= \bk{\pi_0 C_0 + \pi_1 C_1}^{-1}\Delta,
\label{Htilde}
\\
\tilde\lambda(y) &= \tilde H^\top\bk{y-\bar y}
\test 0,
\end{align}
where
\begin{align}
\Delta &= \frac{1}{2}\bk{\Avg{y}_1-\Avg{y}_0},
\\
\bar y &= \frac{1}{2}\bk{\Avg{y}_1+\Avg{y}_0},
\\
C_0 &= \Avg{y y^\top}_0-\Avg{y}_0\Avg{y}_0^\top,
\\
C_1 &= \Avg{y y^\top}_1-\Avg{y}_1\Avg{y}_1^\top,
\end{align}
and the upper error bounds are given by
Eqs.~(\ref{Pebounds})--(\ref{R}).

$\tilde\lambda(y)$ for $P = 1$ is a linear filter with respect to $y$
and similar to the linear filters proposed in
Ref.~\cite{gambetta07}. An advantage of the $\mathcal R$-optimal rule
here is that the filter $\tilde H$ depends only on the first-order moments
$\Delta(k)$ and $\bar y(k)$ and second-order correlations $C_0$ and
$C_1$. All these moments can be simulated or measured directly in an
experiment without the assumptions of continuous time, white Gaussian
noise, and uncorrelated signal and noise made in prior work.  The
calculation of $\tilde H$ is relatively straightforward compared with
the numerical optimization procedure in Ref.~\cite{gambetta07}, while
$\mathcal Q$ and $\tilde{\mathcal R}$ provide theoretical performance
guarantees. The upper bounds may be conservative, and a more precise
comparison of $\mathcal P_e(\tilde H)$ with other linear or nonlinear
filters \cite{gambetta07,danjou,*danjou15,ng} will require further
numerical simulations and experimental tests.

To proceed further, consider the continuous-time limit.  For the
two-level $x \in \{0,1\}$ process with initial value $x(0) = 1$ and
spontaneous decay time $T_1$ studied in Refs.~\cite{gambetta07,ng}, it
is not difficult \cite{gardiner} to show that the mean is
\begin{align}
\Avg{x(t)} = \exp\bk{-\frac{t}{T_1}},
\end{align}
and the covariance function is
\begin{align}
C_x(t,\tau) &\equiv  \Avg{x(t)x(\tau)}-\Avg{x(t)}\Avg{x(\tau)}
\\
&= 
\exp\Bk{-\frac{\max(t,\tau)}{T_1}}-\exp\bk{-\frac{t+\tau}{T_1}}.
\end{align}
For a zero-mean white Gaussian noise with noise power $\Pi$,
\begin{align}
\Avg{y_0(t)y_0(\tau)} &= \Pi\delta(t-\tau).
\end{align}
The test statistic becomes
\begin{align}
\tilde\lambda &= \int_0^T dt \tilde h(t)\Bk{y(t)-\frac{S}{2}\Avg{x(t)}},
\end{align}
and a continuous-time limit of Eq.~(\ref{Htilde}) leads to a
Fredholm integral equation of the second kind \cite{vantrees} given by
\begin{align}
\frac{S}{2}
\Avg{x(t)} &= \Pi \tilde h(t) + \pi_1 S^2\int_0^T d\tau C_x(t,\tau)\tilde h(\tau).
\label{fredholm}
\end{align}
Further analytic simplifications may be possible for $T\to\infty$
using Laplace transform, but a numerical solution of the Fredholm
equation can easily be sought, as it is linear with respect to
$\tilde h_1$ and can be inverted in discrete time using, for example,
the \texttt{mldivide} function in Matlab.

\begin{figure}[htbp]
\centerline{\includegraphics[width=0.48\textwidth]{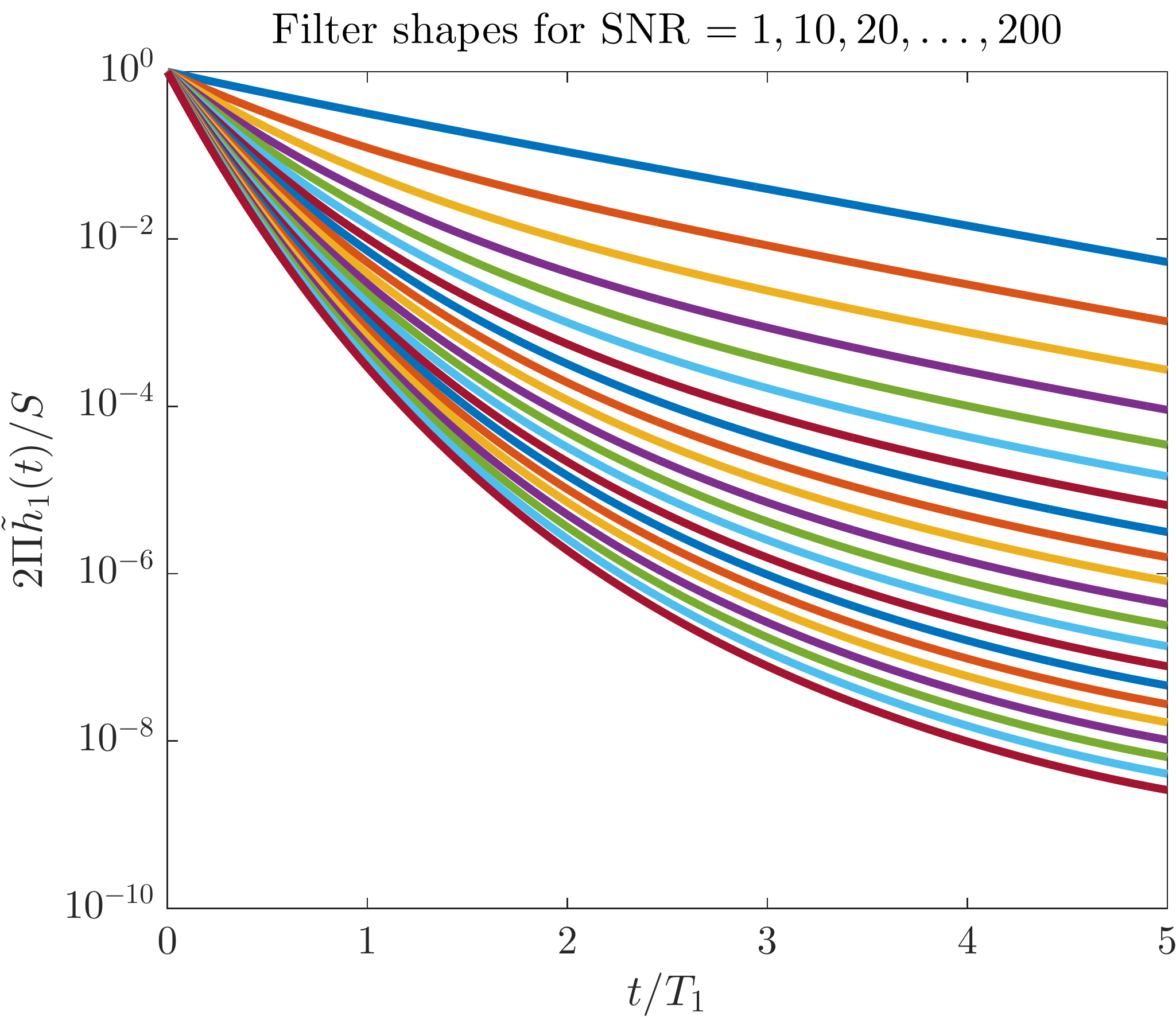}}
\caption{(Color online). The normalized $\mathcal R$-optimal filters
  $2\Pi\tilde h_1(t)/S$ in log scale versus normalized time $t/T_1$
  for different input
  $\textrm{SNR}\equiv S^2 T_1/\Pi = 1,10,20,\dots,200$.
  $\pi_0 = \pi_1 = 1/2$ and $T = 5T_1$ are assumed. The different
  plots can be distinguished by the reducing correlation times for
  increasing SNR.}
\label{matched_filters}
\end{figure}

\begin{figure}[htbp]
\centerline{\includegraphics[width=0.48\textwidth]{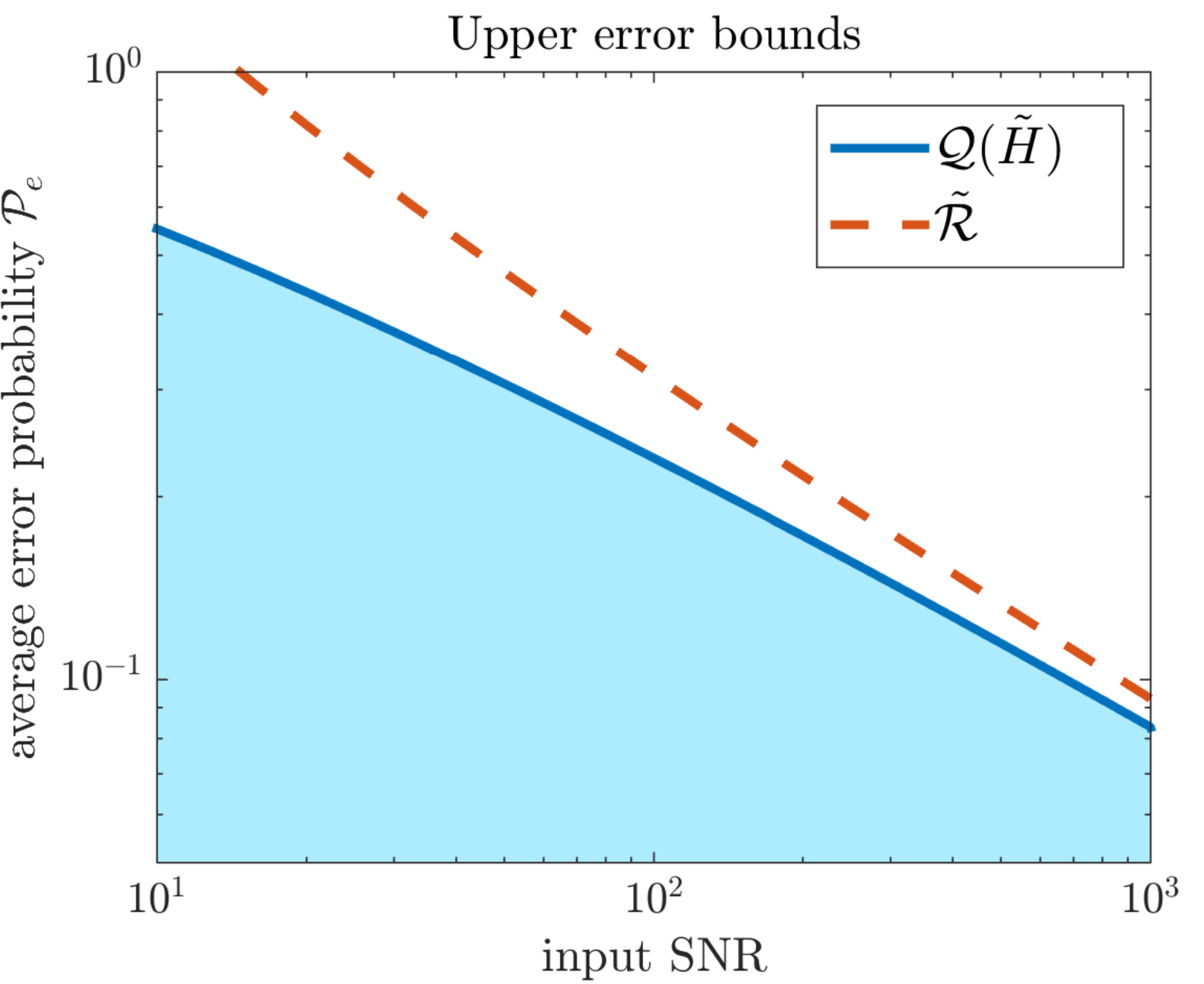}}
\caption{(Color online). Upper bounds $\mathcal Q(\tilde H)$ and
  $\tilde{\mathcal R}$ on the average error probability $\mathcal P_e$
  for the first-order Volterra filter versus input SNR from 10~dB to
  30~dB in log-log scale.  $\pi_0 = \pi_1 = 1/2$ and $T = 5T_1$ are
  assumed.  $\mathcal P_e$ is guaranteed to be in the shaded region
  below the curves.}
\label{upper_bounds}
\end{figure}

Define the input signal-to-noise ratio (SNR) as
$S^2T_1/\Pi$. Fig.~\ref{matched_filters} plots some numerical examples
of the filter for $\pi_0 = \pi_1 = 1/2$ and $T = 5T_1$.  The Matlab
computation of all the filters shown with $\delta t = 0.001 T_1$ takes
seconds to complete on a desktop PC.  Fig.~\ref{upper_bounds} plots
the upper error bounds versus the input SNR. The upper bounds turn out
to be conservative here, as a numerical investigation of
$\mathcal P_e$ later will demonstrate.

The proposed decision rule can be compared with the optimal
likelihood-ratio test (LRT) \cite{vantrees}. For the given problem,
there exists an analytic expression for the log-likelihood ratio given
by \cite{ng}
\begin{align}
\lambda_o(y) &=  \frac{S}{\Pi}\int_0^T d\eta(t) \check x(t)-\frac{S^2}{2\Pi}
\int_0^T dt \check x^2(t),
\label{lambdaO}
\\
d\eta(t) &= y(t) dt,
\\
\check x(t) &= \frac{p_1(t)}{p_0(t)+ p_1(t)},
\\
p_1(t) &= \exp\Bk{\frac{S}{\Pi}\int_0^t d\eta(\tau) 
-t \bk{\frac{S^2}{2\Pi}+\frac{1}{T_1}}},
\\
p_0(t) &= \frac{1}{T_1}\int_0^t d\tau p_1(\tau),
\end{align}
where the $d\eta$ integrals are in the It\={o} sense. The optimal
decision rule is thus
\begin{align}
\lambda_o(y) \test \ln \frac{\pi_0}{\pi_1}.
\end{align}
Although the LRT will achieve the lowest $\mathcal P_e$, the highly
nonlinear dependence of $\lambda_o$ on $y$ makes its exact
implementation difficult in real-time applications or for a large
number of qubits.

\begin{figure}[htbp]
\centerline{\includegraphics[width=0.48\textwidth]{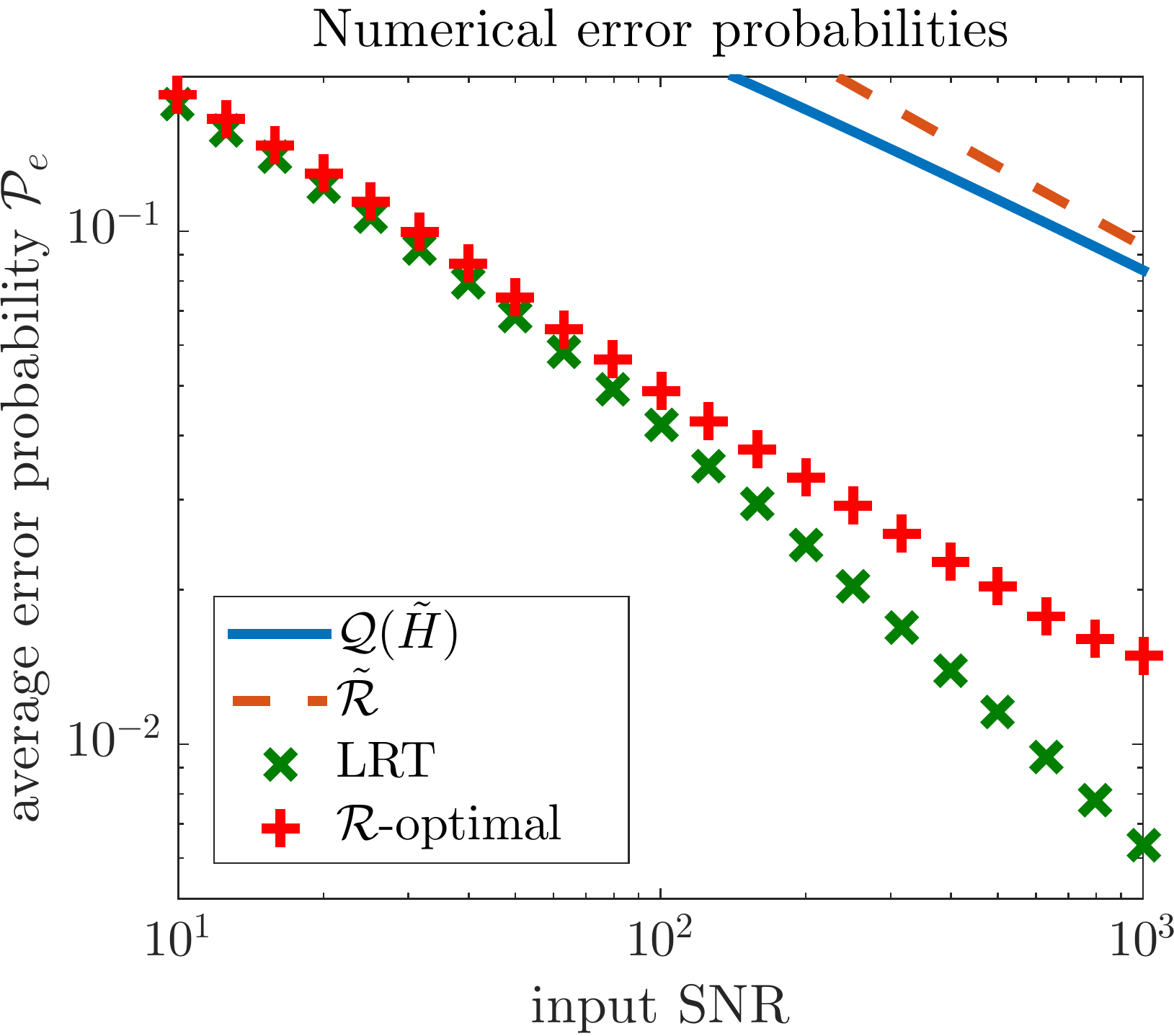}}
\caption{(Color online). Numerically computed average error
  probabilities $\mathcal P_e$ for the $\mathcal R$-optimal rule and
  the likelihood-ratio test (LRT) versus the input SNR from 10~dB to
  30~dB in log-log scale. $\pi_0 = \pi_1 = 1/2$ and $T = 5T_1$ are
  assumed.  Also shown are parts of the upper bounds
  $\mathcal Q(\tilde H)$ and $\tilde{\mathcal R}$ for comparison.}
\label{numerical_Pe}
\end{figure}

The average error probabilities for both the $\mathcal R$-optimal rule
and the LRT are estimated numerically using Monte Carlo simulations
and plotted in Fig.~\ref{numerical_Pe}. The errors are close at lower
input SNR values. Considering the simplicity of the
$\mathcal R$-optimal rule, the divergence at higher SNR is expected
and indeed slight. At the input SNR of $10^3$, $\mathcal P_e$ for LRT
is $6.2\times 10^{-3}$, while that for the $\mathcal R$-optimal rule
is only around a factor of 2 higher at $1.48\times 10^{-2}$. A further
optimization of $\mathcal P_e$ beyond the results shown in
Fig.~\ref{numerical_Pe} can be done by fine-tuning the threshold of
the $\mathcal R$-optimal rule. For example, a numerical search for the
optimal threshold brings its error probability at input SNR $= 10^3$
down to $8.3\times 10^{-3}$.  A higher-order filter is hardly
necessary for the SNRs considered here.

The upper bounds depend only on low-order moments and apply equally to
all problems with the same low-order moments, regardless of their
higher-order statistics. It is not surprising that such indiscriminate
bounds are loose for this particular example, as shown in
Fig.~\ref{numerical_Pe}. What is surprising is the near-optimal
performance of a decision rule based on a loose upper bound. The
log-likelihood ratio is given analytically for the problem considered
here, so one may compare it with the $\mathcal R$-optimal test
statistic to see how the two resemble each other. In general, however,
the log-likelihood ratio is difficult or even impossible to compute if
the full probability models are more complicated or simply
unidentified. The $\mathcal R$-optimal rule requires only low-order
moments to be known, and is hence more convenient to implement in
practice.
\section{Conclusion}
I have proposed the use of Volterra filters for quantum estimation and
detection. The importance of the proposal lies in its promise to solve
many of the practical problems associated with existing optimal
quantum inference techniques, including the curse of dimensionality,
the lack of performance assurances upon approximations, and the need
for a Markovian model.  Beyond the examples of quantum state
tomography and qubit readout discussed in this paper, diverse
applications in quantum information processing
\cite{wiseman_milburn,nielsen,aspelmeyer14}, including cooling
\cite{steck,*steck06}, squeezing \cite{thomsen}, state preparation
\cite{yanagisawa06,*negretti}, metrology
\cite{helstrom,holevo,braginsky,glm2011,smooth,twc,qzzb,*qbzzb,hypothesis,tsang_nair,*tsang_open},
fundamental tests of quantum mechanics
\cite{hypothesis,testing_quantum,chen2013,aspelmeyer14}, and error
correction \cite{ahn,*sarovar,chase08}, are expected to benefit.
Potential extensions of the theory include adaptive, recursive, and
coherent generalizations for feedback control \cite{wiseman_milburn}
and noise cancellation \cite{qnc,*qmfs,*yamamoto14}, filter training
via machine learning \cite{hentschel,*magesan}, robustness analysis,
the use of other performance criteria for improved robustness
\cite{james04} or multi-hypothesis testing
\cite{hypothesis,testing_quantum}, a connection with Shannon
information theory through the relations between filtering errors and
entropic information \cite{barchielli_lupieri,*mismatch}, and a study
of fundamental uncertainty relations in conjunction with quantum lower
error bounds
\cite{helstrom,holevo,glm2011,twc,tsang_nair,*tsang_open,qzzb,*qbzzb}.

\section*{Acknowledgments}
This work is supported by the Singapore National Research Foundation
under NRF Grant No.~NRF-NRFF2011-07.

%



\end{document}